**Field-dependent low-field enhancement in effective paramagnetic moment with nano-scaled $Co_3O_4$**

Weimeng Chen[1], Chinping Chen[1a], and Lin Guo[2b]

[1] Department of Physics, Peking University, Beijing, 100871, People's Republic of China

[2] School of Chemistry and Environment, Beijing University of Aeronautics and Astronautics, Beijing, 100191, People's Republic of China

Abstract:

Paramagnetic (PM) properties of columnar cobaltosic oxide ($Co_3O_4$) nanoparticles, about 75 nm in diameter, have been investigated by magnetization measurements at $T > T_N = 39$ K. In zero or low applied field, the effective PM moment per formula unit (FU), $\mu_{eff}$, enhances significantly from the bulk value of 4.14 $\mu_B$/FU. It decreases asymptotically from 5.96 $\mu_B$/FU at $H_{app} = 50$ Oe down to 4.21 $\mu_B$/FU as the applied field increases to $H_{app} = 10$ kOe. The field dependent PM properties are explained by a structural inversion, from the normal spinel (spin-only moment ~ 3.9 $\mu_B$/FU) to the inverse spinel structure (spin-only moment ~ 8.8 $\mu_B$/FU). The structural inversion is reversible with the variation of the applied field. The lattice structure becomes almost 100% normal spinel in the high field exceeding 10 kOe, as revealed by the magnitude of the effective PM moment. The reversible, field dependent structure inversion is an important property with promising application potential. It is interesting for the future investigations.

[a] E-mail: cpchen@pku.edu.cn, Phone: +86-10-62751751
[b] E-mail: guolin@buaa.edu.cn

## I. Introduction

Cobaltosic oxide ($Co_3O_4$) has shown excellent electrochemical cycling properties and is under intensive consideration as an electrode material for advanced lithium ion batteries [1-4], owing to the electrochemical reaction mechanism of electron transfer between the $Co_3O_4$ and Li [4]. Also, it has received increasing attentions recently because of its catalytic properties of CO oxidation even at temperature much below room temperature, at about – 77 $^o$C [5, 6]. The $Co^{3+}$ ions on the octahedral sites play the active role for the catalysis of CO oxidation, while the $Co^{2+}$ ions on the tetrahedral sites are inactive. Because of its multiple functionalities with promising application potentials, a great effort has been made in synthesizing different sizes and morphologies of $Co_3O_4$ nanoparticles and in studying the properties, especially the electric and magnetic properties. Bulk $Co_3O_4$ has a normal cubic spinel structure, $AB_2O_4$, with the spin moments of the $Co^{2+}$ ions (4.14 $\mu_B$) at the tetrahedral A-sites and of the $Co^{3+}$ ions (0 $\mu_B$) at the octahedral B-sites [7, 8]. For magnetic properties, $Co_3O_4$ is antiferromagnetic (AFM) with each $Co^{2+}$ ion in the A-site having four neighboring $Co^{2+}$ ions of opposite spins. The Néel temperature of $Co_3O_4$ bulk crystal is known as, $T_N \sim 40$ K [7], and it reduces with the particle size ascribed to the finite size effect [9]. The crystal field of the octahedral symmetry in the B-sites splits up the 3$d$ orbitals into three lower energy $t_{2g}$ levels and two higher energy $e_g$ levels with a large band gap in between, ~ 2.0 eV ( ~ 19000 cm$^{-1}$) [7, 10]. Therefore, the $Co^{3+}$ (3$d^6$) ions at the B-sites is in the low spin (LS) state, $S = 0$, and it is without magnetic moment. In this state, the 6 $d$-electrons fill up the three lower energy $t_{2g}$ levels since the Hund's rule gives in to the large band gap. On the other hand, the tetrahedral symmetry in the A-sites makes the 3d orbitals split into three higher energy $t_{2g}$ and two lower energy $e_g$ levels, with a smaller band gap of about 0.4 eV ( 3700 cm$^{-1}$) [7]. The Co$^{2+}$ (3$d^7$) ions at the A-sites are in the high spin (HS) state with $S = 3/2$, i.e., four d-electrons fill up the two lower energy $e_g$ orbitals while three others half-fill the $t_{2g}$ orbitals [8]. Therefore, the observed magnetism with Co$_3$O$_4$ normally arises from the A-site Co$^{2+}$ ions, while the B-site Co$^{3+}$ ions make essentially no contribution. The effective magnetic moment, $\mu_{\text{eff}}$, of bulk Co$_3$O$_4$ is determined experimentally as 4.14 $\mu_B$ [7], which is larger than the spin-only moment, $2[S(S+1)]^{1/2} = 3.87$ $\mu_B$, calculated in theory using the gyromagnetic ratio, $g = 2$ for the free Co$^{2+}$ ions. This indicates that there is a slight contribution from the orbital moment [7].

By reducing the size of Co$_3$O$_4$ to nanoscale, some of the interesting magnetic properties have been investigated [9, 11-17]. In particular, a weak FM behavior has been observed at $T < T_N$ for Co$_3$O$_4$ nanoparticles [17]. It is explained that a noticeable part of the normal spinel structure is inverted to the inverse spinel structure by the "site-exchange" of the Co$^{2+}$ ions on the octahedral B-sites and the Co$^{3+}$ on the tetrahedral A-sites. This occurs probably arising from a charge transfer, which has a much lower energy barrier than does a direct swap of the ions. The effective moment is enhanced by about 25% from the bulk value. For the Co$^{3+}$ ions at the tetrahedral A-sites of the inverse spinel lattice, the energy gap separating the lower energy $e_g$ and the higher energy $t_{2g}$ orbitals reduces from 0.4 eV by about 8% [7]. The Hund's rule thus gives in. This energetically favors the high spin (HS) state, $S = 2$ for the Co$^{3+}$ ions, for which only three *d*-electrons occupy the two lower energy $e_g$ orbitals, leaving one unpaired electron, and the other three electrons half-fill the three $t_{2g}$ orbitals of higher energy. For the $Co^{2+}$ ions at the B-sites with the inverse spinel structure, on the other hand, the high spin (HS) state, $S = 3/2$, is also energetically favored over the low spin (LS) state, $S = 1/2$. This is because that the energy gap due to the octahedral crystal field is much reduced from about 2 eV down to ~ 1.1 eV ( ~ 10200 $cm^{-1}$) [7]. With the variation of magnetic/crystal structure mentioned above, a corresponding fraction of the AFM ordering thus becomes frustrated. The effective magnetic moment per formula unit (FU), $\mu_{eff}$, increases because of the transformation of lattice structure, i.e. $Co^{2+}$ (A sites, $S = 3/2$) + $Co^{3+}$ (B sites, $S = 0$) for the spinel structure becomes $Co^{3+}$ (A sites, $S = 2$) + $Co^{2+}$ (B sites, $S = 3/2$) for the inverse spinel structure [7, 8]. The structural transition which gives rise to a spatially inhomogeneous distribution of magnetic moments and incommensurate AFM ordering has been reported even for a $Co_3O_4$ polycrystalline sample by a positive muon spin rotation and relaxation method. This is also ascribed to the spin state changes of the lattice sites arising from charge transfer [18].

For the spinel crystal structure, the scenario of electron transfer between different elemental ions on A and B sites has also been reported, e.g., with $CoFe_2O_4$ [19] and $Fe_3O_4$ [20] by magneto-optical effects. An enhancement of magnetization observed for $ZnCr_2O_4$ nanocrystals is attributed to a structural transformation [21]. It is even reported that partly structural change occurs with the powder sample of nanocrystalline $ZnFe_2O_4$ by mechanical activation of a ball-milling process which causes a direct exchange of elemental atoms on different cation sites [22-24]. In addition, the particle-size-dependent magnetization has been reported with 8 nm $CdFe_2O_4$ particles [25]. In fact, it is known that a direct "site-exchange" between the A and B sites atoms easily occurs at high temperature because of thermal activation, which gives rise to a mixing structure with a different degree of inversion depending on the synthesis temperature. Obviously, it favors the mechanism by the charge transfer at low temperature, rather than by the direct exchange of ions between the lattice sites.

In this paper, we investigate systematically the paramagnetic (PM) behavior of 75 nm columnar $Co_3O_4$ nanoparticles, both at $T > T_N$ and $T < T_N$. The effective magnetic moment per FU, $\mu_{eff}(H_{app})$ is shown to depend on the applied field, $H_{app}$. At $T > T_N$ in the PM phase. It is determined as $\mu_{eff} \sim 5.96$ $\mu_B$/FU at $H_{app} = 50$ Oe by the analysis of Curie-Weiss law on the $M(T)$ measurements, and the magnitude reduces asymptotically toward the bulk value of $4.14\mu_B$/FU with the increasing applied field reaching $H_{app} = 10$ kOe. In addition, by the $M(H)$ measurements at $T < T_N$, the enhancement of the effective moments is observed also at the presence of the AFM $Co_3O_4$ phase. The evidences indicate that there already exists a fraction of inverse spinel structure along with the normal spinel structure in $H_{app} = 0$ Oe and in low field. Therefore, it exhibits the magnetic properties of HS states. Under the applied field, however, part of the inverse spinel structure undergoes a structural transformation, inverted back to the spinel structure. The field dependent inversion ratio, $R(H_{app})$, which describes the fraction of the inverse spinel structure over the whole crystal lattice is determined as $R \sim 42.6 \%$ at $H_{app} = 50$ Oe.

The synthesis is by a wet chemical solution method. Detailed synthesis procedure and the characterizations of the sample are published in a previous report [9]. The sample under present investigation is in the form of a columnar structure with a diameter of 75 nm. The Néel temperature is determined as $T_N = 39$ K, slightly reduced from the bulk value due to the finite size effect [9].

**II. Magnetic measurements**

Magnetization of the powder sample, ~1.93 mg, has been measured using a SQUID magnetometer (Quantum Design). A series of zero-field-cooled (ZFC) temperature dependent magnetization, $M_{ZFC}(T)$, curves are recorded in the applied field, $H_{app} = 50$, 90, 150, 250, 600 Oe, 2 kOe, and 10 kOe, in the increasing temperature after the sample is cooled from room temperature down to 5 K without the applied field. The magnetic susceptibility, $\chi = M_{ZFC}(T)/H_{app}$, is then calculated from the data. In addition to the $M_{ZFC}(T)$ measurements, $M(H)$ measurements at $T = 10, 30, 50, 100$ and 250 K have also been performed.

a) $M(T)$ measurements

The temperature dependence of ZFC magnetization, $M_{ZFC}(T)$ is shown in Fig. 1. The magnetization curves are presented in a log-Y scale. Unlike the freezing temperature of a spin glass state [26] or the blocking temperature of ferri/ferromagnetic materials [27-29], the peak position does not show any variation with the increasing field. This is a characteristic feature for an AFM transition. The Néel temperature occurs at $T_N$ = 39.1 K. It is slightly reduced from the bulk value of 40 K, attributed to the finite size effect [9].

The inverse of the susceptibility, $1/\chi(H_{app})$, exhibits a well-behaved linear property from 300 K down to 50 K under various applied field, $H_{app}$, as shown in Fig. 2. The solid lines are for the fitting results to the data points by the Curie-Weiss law, $\chi = \dfrac{N\mu_0 \mu_{eff}^2}{3k_B(T-T_\theta)}$, where $N$ = 2.5×10$^{21}$ is the value calculated for the number of formula unit (FU) included in 1 g of $Co_3O_4$, $\mu_{eff}$ is the effective magnetic moment per FU, $T_\theta$ is a parameter for the Curie-Weiss temperature, and $k_B$ is the Boltzmann constant. This indicates that the mean field approximation is valid irrespective of the magnitude of the applied field. From the fitting analysis, $T_\theta$ is determined from the extrapolation of the solid line to $1/\chi = 0$. So, the effective moment per FU at $H_{app}$ = 50 Oe is determined as, $\mu_{eff}$ = 5.96 $\mu_B$/FU and the Curie-Weiss temperature as, $T_\theta$ = - 201 K. With $T_\theta < 0$, it indicates that the exchange interaction between the PM magnetic moments is of AFM nature. Unlike the Néel temperature which is field independent as shown in Fig. 1, the Curie-Weiss temperature goes down with the increasing field. It implies the variation of the effective mean field, which reflects the magnitude of the effective PM moment.

Fig. 3 shows the field dependence of the effective magnetic moment, $\mu_{eff}(H_{app})$ in units of $\mu_B$/FU, and the Curie-Weiss temperature, $T_\theta$, estimated from the inverse susceptibility of the $M_{ZFC}(T)$ curves. The effective magnetic moment decreases asymptotically from 5.96 to 4.21 $\mu_B$/FU as the applied field increases from 50 Oe to 10 kOe. Correspondingly, the magnitude of Curie-Weiss temperature decreases with the increasing applied field, from - 201 K in 50 Oe to - 77 K in 10 kOe.

b) $M(H)$ measurements

The field dependent $M(H)$ curves measured at 10, 30, 50 100, and 250 K, are shown in Fig. 4. All $M(H)$ curves of the sample exhibit PM-like linear behavior, however, with a slight nonlinearity in the low field region near $H = 0$ Oe. The slope of the $M(H)$ curves increases locally as the applied field approaches 0 Oe. Two of the $M(H)$ curves, one at $T = 30 < T_N$ and the other at $T = 250 \text{ K} > T_N$, are shown in the insets. The straight lines are obtained by a linear fitting to the lower half branches of the two $M(H)$ data sets from $H = - 10$ to $- 45$ kOe, respectively, and moved upwards in parallel to go through the origin. It clearly reveals the nonlinearity in the region around $H = 0$ Oe. The inset also shows that the $M(H)$ curves are fully reversible at the sweeping of the applied field, as expected for the PM properties already observed by the $M_{ZFC}(T)$ measurements. Furthermore, as the applied field increasing over 4 kOe, the linearity manifests itself. This is in consistent with the results shown in Fig. 3 that the variation of the effective moment approaches its high field value at the applied field of several kOes. Since the susceptibility corresponds to the slope of the $M(H)$ curves, $\chi(H_{app}) = dM(H)/dH|_{H = H_{app}}$, it further confirms that the effective magnetic moment of the sample increases as the applied field approaches $H = 0$ Oe.

IV. Analysis and discussion

The observed enhancement of the effective PM moments with the nanoscaled $Co_3O_4$ sample is attributable to the fact that a part of the crystal is in the inverse spinel instead of the spinel structure at zero applied field. Because of the structural inversion, a part of the $Co^{2+}$ ions at the tetrahedral A-sites thus becomes $Co^{3+}$, while an equal number of $Co^{3+}$ ions at the octahedral B-sites become $Co^{3+}$ ions. As a result, the effective magnetic moment are enhanced because of the change in the spin states of Co ions, i.e., the inversion from the spinel structure, $Co^{2+}$(A sites, $S = 3/2$) + $Co^{3+}$(B sites, $S = 0$) to the inverse spinel structure, $Co^{3+}$(A sites, $S = 2$) + $Co^{2+}$(B sites, $S = 3/2$) leads to the increment of the spin-only moments. For a fully inverse spinel structure, the spin-only moment of $Co_3O_4$ is calculated as 8.77 $\mu_B$/FU. In our experiment, the enhanced effective magnetic moment, ~ 5.96 $\mu_B$/FU at $H = 50$ Oe, is significantly larger than the value, ~ 4.14 $\mu_B$/FU of the bulk $Co_3O_4$. The field dependent inversion ratio is then estimated as $R(H_{app})$, ~ 42.6% at $H_{app} = 50$ Oe, i.e., about 42.6% of the sample is of the inverse spinel structure in the low field or zero applied field. As the applied field increases to 10 kOe, the effective moment decreases to 4.21 $\mu_B$/FU, close to the bulk value. It is noted that spin unpairing transitions of the octahedral $Co^{3+}$ ions is possible for the $d$-electrons in the lower energy $t_{2g}$ orbitals to the higher energy $e_g$ orbitals as the Hund's rule gives in to the thermal excitation, i.e., $Co^{3+}$ (LS, $S=0$) becomes $Co^{3+}$ (HS, $S=1$ or 2). However, the temperature range for this to occur is from 600 to 1200 K [30], much higher than the temperature range of this work. This scenario is therefore unlikely for the present experiment.

The AFM transition temperature, $T_N$ ~ 39.1 K, is independent of the applied field

[9], and it reflects only the part of the moments with the spinel structure, which participate in the AFM ordering, irrespectable of the PM moments with the inverse spinel structure. However, the Curie-Weiss temperature, $T_\theta$, determined from the fitting analysis of the $M_{ZFC}(T)$ curves is a result of the mean field approximation accounting for the contribution of even the PM moments with the inverse spinel structure. For the $Co_3O_4$ nanoparticles, the mean field theory is adequate because of the well-behaved Curie-Weiss properties shown in Fig. 2 and is also supported by the results of finite size effect with the shift exponent, $\lambda = 1.1 \pm 0.2$ [9]. Thus, the variation of the effective moments with the Co ions is consistent with the result that $T_\theta$ changes with the applied field, as shown in Fig. 3. The ratio of $T_\theta$ over $T_N$ is defined as the geometric frustration parameter $f = T_\theta / T_N$ [31], which reflects the degree of frustration in the AFM ordering state. It decreases with the increasing applied field, from 5.1 in $H_{app} = 50$ Oe to 2.0 in 10 k Oe. The geometrically frustrated effect is particularly strong for spins with AFM interactions on the three-dimensional network of corner-sharing tetrahedral [32]. Recently, this phenomenon has been demonstrated for the spinel structure by the experiments of neutron scattering investigation on cubic spinel $ZnCr_2O_4$ [33] and electron-spin resonance of $FeAl_2O_4$, $CoAl_2O_4$ and $MnAl_2O_4$ [31]. The value of $f$ is large in low applied field and decreases with the increasing applied field, owing to the partial inversion effect in low field.

The property of the field-induced structural change indicates the strong coupling of the magnetic atoms with the magnetic field. Usually, the structure transformation induced by the magnetic field occurs at several tens of kOes [34, 35]. For example,

Gd$_5$Ge$_4$ changes from AFM to a collinear ferromagnetism below 30 K when the field exceeds 10 kOe [34]. However, these phenomena rather differ from our results on Co$_3$O$_4$. The inversion ratio and the effective magnetic moment of Co$_3$O$_4$ are sensitive to the magnetic field in the low field region. Several tens of Oe can make the paramagnetic properties change appreciably. Perhaps, this is because the inversion ratio, $f$, has already approached zero at the field below 10 kOe, i.e., almost all of the inverse spinel structure has almost inverted to the spinel structure in the low field region.

Generally, the structural inversion for a sample of spinel structure becomes more significant as the particle size reduces by the reason of increasing surface-to-volume ratio, and the surface changes or modifies the crystal field symmetry easily. It results in an increasing of the effective magnetic moment and magnetic susceptibility with the reducing particle size [21, 24, 25]. In addition, the other factors, such as the synthesis process, postprocessing history, etc, also play an important role in determining the inversion ratio. For example, Atif *et al*. observed that the inversion ratio of ZnFe$_2$O$_4$ with the same particle size obtained by using sol-gel method in basic medium is larger than that obtained by using sol-gel method in acidic medium [24]. Hamdeh *et al*. reported that ball-milling increases the inversion ratio from 0.21 of the original 10 nm ZnFe$_2$O$_4$ product to a value of 0.55 [23]. Additionally, the synthesis temperature also influences the inversion ratio of ions between A and B sites [31].

V. Conclusion

In summary, we investigated systematically the magnetic properties of columnar $Co_3O_4$ nanoparticles, about 75 nm in diameter. The PM properties appear to be well described by the Curie-Weiss law, indicating that the mean field theory is appropriate for this material. The effective PM moment is field dependent, estimated as 5.96 $\mu_B$/FU in $H_{app}$ = 50 Oe and reduces down to 4.21 $\mu_B$/FU as $H_{app}$ reaches 10 kOe. In addition, the Curie-Weiss temperature reduces with the applied field, while the Néel temperature remains unchanged. The enhancement of the effective PM moment is ascribed to the inversion of the lattice from the spinel to the inverse spinel structure at zero or low field. Consequently, part of the Co cations change from the LS to HS states i.e. $Co^{2+}$ (A sites, $S$ = 3/2) + $Co^{3+}$ (B sites, $S$ = 0) for the spinel structure becomes $Co^{3+}$ (A sites, $S$ = 2) + $Co^{2+}$ (B sites, $S$ = 3/2) for the inverse spinel structure. The inversion ratio is estimated as 42.6 % in the applied field of 50 Oe. Although without a direct experimental evidence at this stage, we believe that the field-dependent structural inversion is most likely attributed to the electron hopping between the Co cations, rather than a direct "exchange" of the Co ions.

Figures

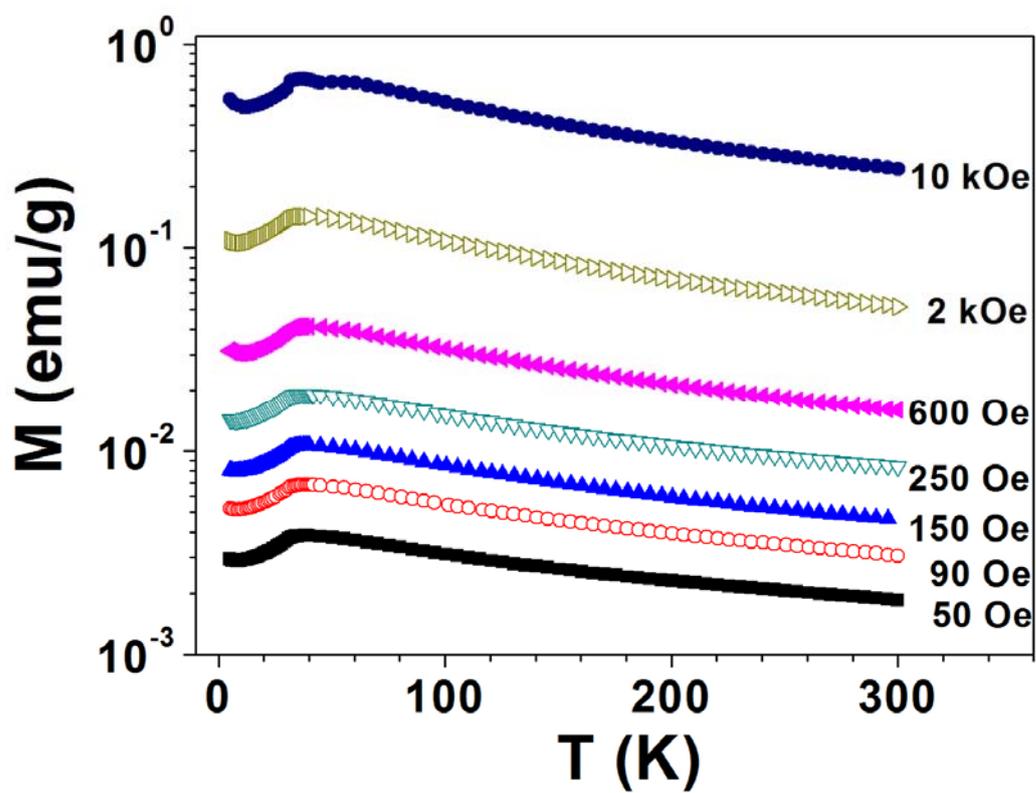

Fig. 1 Zero-field-cooling (ZFC) $M(T)$ curves measured in various applied fields from 50 Oe to 10 kOe with the temperature increasing from 5 to 300 K. The magnetization curves are presented in the log-Y scale.

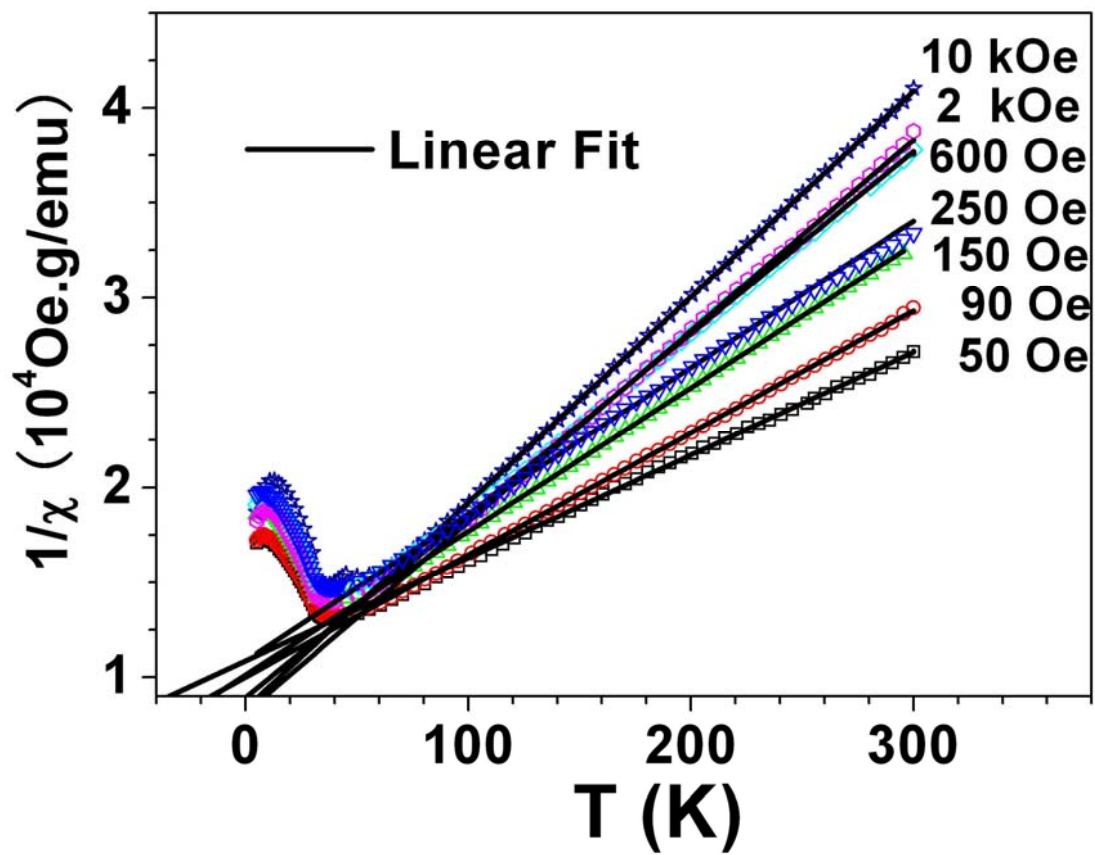

Fig. 2 Inverse susceptibilities, $1/\chi$, determined from the ZFC curves. The solid lines correspond to the linear fitting result of the data, at $T > T_N$ by the Curie-Weiss law.

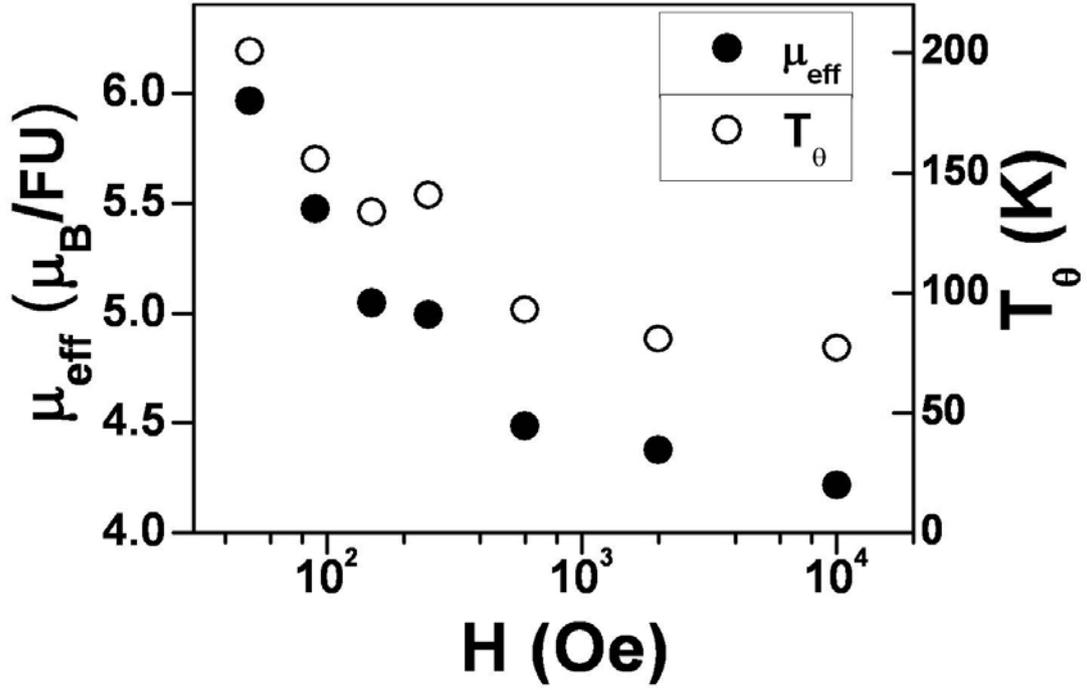

Fig. 3 Field dependent effective magnetic moment per formula unit, $\mu_{eff}(\mu_B/\text{FU})$, and paramagnetic Curie-Weiss temperature, $T_\theta$. The values of $\mu_{eff}$ and $T_\theta$ are calculated from the fitting results of the inverse magnetic susceptibility, $1/\chi$, of the $M_{ZFC}(T)$ curves.

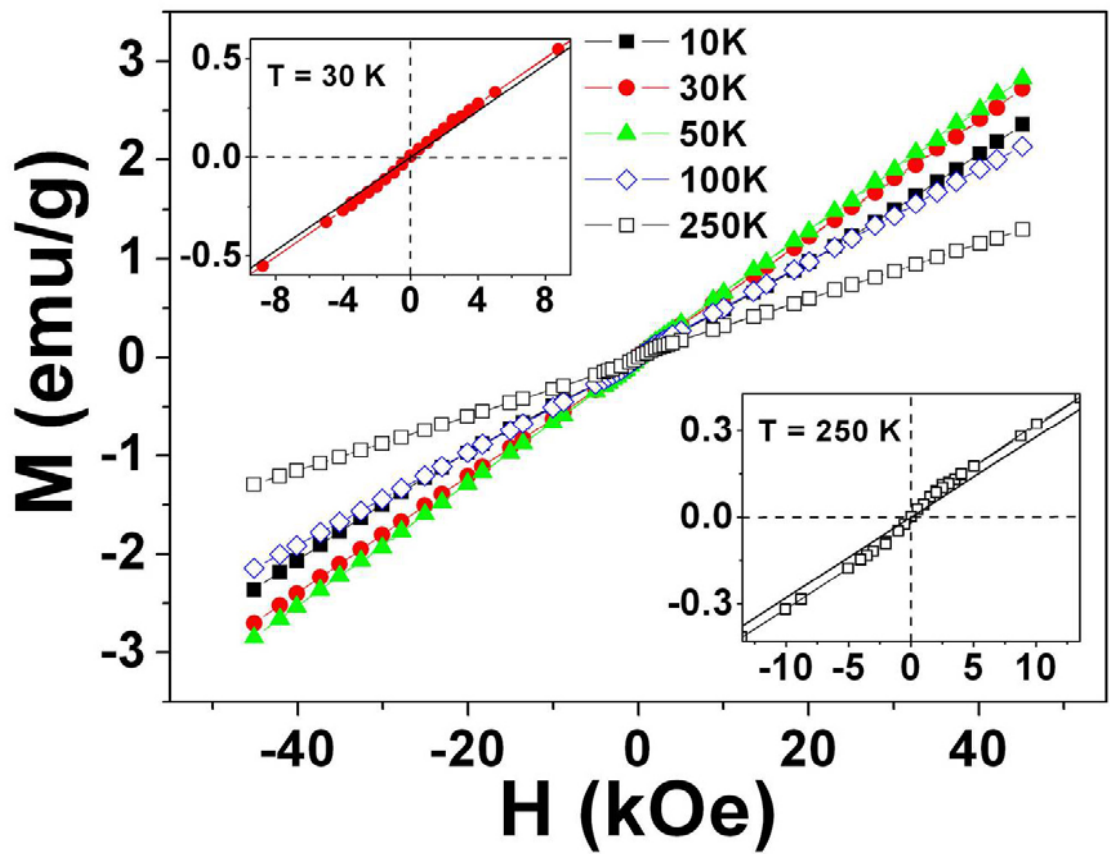

Fig. 4 Field-dependent *M*(*H*) curves. The insets selectively reveal the low field regions of the *M*(*H*) curves measured at *T* = 30 and 250 K. The linear lines going through the origin in the insets is parallel to the lines determined from the linear fittings to the lower half branches of the *M*(*H*) curves, with *H* ranging from -10 kOe to -45 kOe for the fitting.